 \documentclass[final,5p,times,twocolumn]{elsarticle}
\newcommand{\beq}{\begin{equation}}
\newcommand{\eeq}{\end{equation}}
\usepackage{amssymb}

\journal{Physica C}

\begin{document}

\begin{frontmatter}


\title{Why non-superconducting metallic elements become superconducting under high pressure}


\author{J.E. Hirsch and J.J. Hamlin}

\address{Department of Physics, University of California, San Diego, La Jolla, CA 92093-0319}

\begin{abstract}
We predict that simple metals and early transition metals that become superconducting under high pressures will show a change in sign of their Hall coefficient from negative to positive under pressure.
If verified, this will strongly suggest that hole carriers play a fundamental role in `conventional' superconductivity, as  predicted by the theory of hole superconductivity.
\end{abstract}


\end{frontmatter}


Simple and noble metals are not superconducting at ambient pressure, nor are the early and late transition metals. These elements have in common
that their Hall coefficient ($R_H$) at ambient pressure is negative, unlike most other metallic elements that have positive $R_H$ and do become superconducting at ambient pressure\cite{chapnik}. Here we propose that the unexpected and sometimes remarkably high $T_c$  superconductivity recently discovered in simple metals and early transition metals under high 
pressure\cite{pressure} (see Table I) {\it is due to the fact that these metals acquire hole carriers under high pressure}, that come about through deformations of the lattice structure, as discussed below. We predict that this should give rise
to a change in sign of  the Hall coefficient from negative to positive with increasing pressure. The Hall coefficient of these
metals under pressure has not yet been measured. Because late transition metals and noble metals 
are much less compressible, they cannot be pressurized into becoming superconducting in the same way, and their $R_H$ should remain negative under pressure.

The sign of $R_H$ of elements and compounds at ambient pressure shows a very strong correlation with the existence
($R_H>0$)  and nonexistence ($R_H<0$) of superconductivity\cite{correlations,chapnik}. This was noted long ago and led early researchers (pre-BCS)  to propose that  hole carriers
are essential for superconductivity to occur\cite{holes}. This correlation was further strengthened by the finding   that $both$ hole-doped and electron-doped\cite{edoped} cuprates show dominance of hole carrier transport in the
regime where they are superconducting, as does $MgB_2$. The situation for the pnictides is still unclear. Furthermore, very recently, hole-doped semiconductors have been found to become 
superconducting\cite{holesemicond}. Conventional BCS  theory does not indicate any preference for hole transport over electron transport for superconductivity to occur,   while 
the theory of hole superconductivity  does\cite{hsc}.

\begin{table}
\caption{ 
  Non-superconducting simple and early transition metal elements that become superconducting under pressure. Maximum $T_c$ and corresponding pressure $P$ is given, as well as the 
  Hall coefficient $R_H$ at ambient pressure when known. The Hall coefficient at high pressure $R_H(P)$ has not yet been measured. 
   }
\begin{tabular}{l || c | c  | c | c  }
Element & $T_c $ (K)  & P  (GPa)  & $R_H $   &  $R_H$(P) \cr
 &  &  & $  (10^{-11}m^3/C)$   &   \cr
 \hline
 Li & 20 & 48 &  $-150$   & $>$0 predicted \cr
Cs & 1.3 & 12  &  $-71 $   & $>$0 predicted  \cr
Ca & 25 & 161 &  $-18$   & $>$0 predicted \cr
Sc & 19.6 & 106  &  $-3$   & $>$0 predicted \cr
Y & 19.5 & 115  &  $-10$   & $>$0 predicted \cr
  \hline

 \end{tabular}

\end{table}

Consider a simple metal with one valence electron per atom such as $Li$. The first Brillouin zone will be half-filled (Fig. 1(a)) and the carriers of electric current will not have hole-like
character. Hence the metal will not be a superconductor according to the theory of hole superconductivity. $If$ under pressure the lattice were to deform so that in the new structure there are
two atoms per unit cell, a new set of Bragg planes appears, as shown schematically in Fig. 1(b), and there will be small pockets of hole-like carriers.
These carriers will drive the system superconducting according to  the theory of hole superconductivity,   as has been shown for both one-band\cite{hsc} and
two-band\cite{twoband} models. Another consequence of the transition shown in Fig. 1 is that the size of the Fermi surface will decrease substantially leading to 
higher resistivity and possibly even semiconducting or insulating behavior if the hole and electron pockets completely disappear.
\begin{figure}
\resizebox{8.0cm}{!}{\includegraphics[width=7cm]{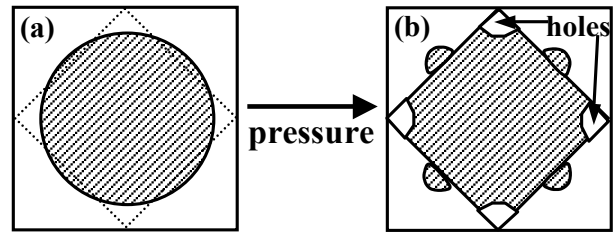}}
  \caption{(a) At ambient pressure, the structure is simple and the first Brillouin zone is approximately half full. (b) Under pressure, the lattice deforms,
  new Bragg planes appear and   the first Brillouin zone becomes almost full.  }
\end{figure}
Furthermore, for a given structure application of pressure will decrease the interatomic distance and hence enhance the carrier hopping amplitude between sites, which
should strongly increase   $T_c$ according to the theory of hole superconductivity\cite{pressure2}.

There is much experimental evidence that simple and early transition metals adopt low symmetry structures with several atoms per unit cell under pressure\cite{hr}, with increased 
electrical resistivity and decreased metallic reflectance. For example, $Li$  shows an almost monotonic resistivity increase of more than 7 orders of magnitude between
ambient pressure and 100GPa, as it undergoes several structural phase transitions and eventually  turns semiconducting and of dark color above 80 GPa\cite{li}.
Similarly $Ca$ shows a nearly monotonic rise in resistivity of several orders of magnitued till it reaches its highest $T_c=25K$ at $161 GPa$\cite{ca}.

Undoubtedly the detailed behavior in each case is very complicated. However, generically we can interpret this behavior as arising from transitions of the type shown in Fig. 1\cite{hr}. Within the Drude expression 
for the resistivity $
\rho=m^*/ne^2\tau$ the resistivity will increase as the number of carriers $n$ decreases due to the extra Bragg planes 
and the average effective mass $m^*$ will increase due to higher `dressing' of the hole carriers\cite{hsc}.

Degtyareva\cite{hr} has argued persuasively that the behavior of these metals under pressure is determined by the Fermi sphere-Brillouin zone (BZ) interaction. Under pressure the electronic 
contribution to the energy dominates over the electrostatic ionic contribution (the former goes as $V^{-2/3}$, the latter as $V^{-1/3}$, with $V$ the volume), 
and the electronic energy is lowered by the appearance of new Bragg planes near the Fermi surface
as the system adopts structures of lower symmetry. This will also lead to the appearance of hole carriers.

A well-studied analogous system is the $Cu-Zn$ alloy, that undergoes a sequence of transitions (Hume-Rothery $\alpha$, $\beta$, $\gamma$ phases) as the $Zn$ concentration
increases. As first proposed by Mott and Jones\cite{mj} the system lowers its energy by creating new Bragg planes through lattice deformation that accommodate increasingly well the
free electron Fermi surface, leading to increasingly higher filling of the first BZ by electrons. For the $\alpha$, $\beta$ and  $\gamma$ phase boundaries the filling of the BZ is
0.62, 0.75 and 0.93 respectively. The $\gamma$ phase has the highest resistivity. Hall coefficient measurements\cite{hall} show a
change in sign from negative to positive as the $Zn$ concentration increases, and hole conduction dominates for high $Zn$ concentration ($\gamma $ phase).

Degtyareva\cite{hr} points out that the sequence of phase transitions and increasing resistivity in $Li$ and other alkali metals with increasing pressure can be interpreted with the Hume-Rothery concept.
For example, above $40GPa$ (where $T_c$ is maximum) the $Li$ structure develops 24 new Bragg planes giving rise to a highly symmetric polyhedron wherein the 
free electron Fermi sphere occupies $90\%$ of the volume. Hence, {\it the system becomes a hole conductor with 0.1 holes per unit cell}. 
Degtyareva proposes that the same physics takes place in alkali metals such as $Ca$, $Ba$ and $Sr$, and we argue that the same is likely to be true for the
early transition metals $Sc$ and  $Y$; their structures under pressure have not yet been 
completely characterized.

We propose that the essential physics can be understood from a simple one-dimensional model, as shown in Fig. 2\cite{dimer}. A nearly half-filled chain
opens up a gap near the Fermi level upon dimerization, and the lower band becomes nearly full, i.e. carriers become 
hole-like.  The wavefunction of an electron at the Fermi level changes sign in hopping to the neighboring unit cell as shown in Fig. 2(a). The correlated hopping interaction\cite{hsc}
\beq
\Delta t  (c_{i \sigma}^\dagger c_{j\sigma}+h.c.)  (n_{i,-\sigma}+n_{j,-\sigma})
\eeq
becomes attractive because $<c_{i \sigma}^\dagger c_{j\sigma}>$ is negative for carriers near the top of the band, and drives the system superconducting.
Furthermore the interaction $\Delta t$ increases as the interatomic distance decreases and drives $T_c$ higher\cite{pressure2}.

In summary, we propose that high pressure gives rise to superconductivity because new Bragg planes in low symmetry high pressure phases give rise to hole-like carriers that
drive the system superconducting\cite{hsc}. Within a given structure, higher pressure reduces the interatomic distances and
increases $\Delta t$ and $T_c$\cite{hsc,pressure2}. The external pressure counteracts the internal pressure in these structures arising from high occupation of {\it antibonding states},
which renders these structures unstable at ambient pressure.
If our explanation is correct, the Hall coefficient    should   go from negative to positive with increasing pressure  in the vicinity where superconductivity appears.

 \begin{figure}
\resizebox{8cm}{!}{\includegraphics[width=6cm]{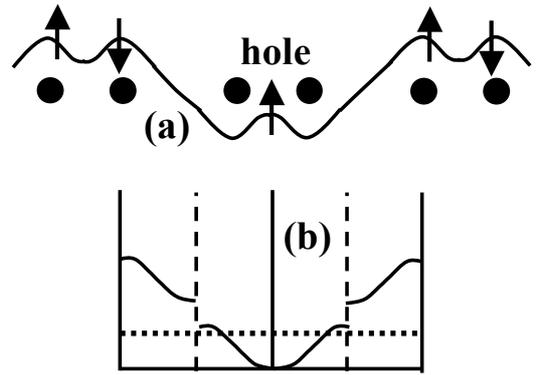}}
  \caption{  When the lattice distorts, the band that was nearly half full becomes nearly full. The top part of the figure shows schematically the wave function in real space
  for a state at the Fermi energy. There is approximately one electron per site. The wavefunction changes sign between unit cells. The bottom part of the figure shows the band in k-space in the extended
  zone scheme and the dotted line indicates the position of the Fermi energy.  }
\end{figure}



\end{document}